\documentclass[doc]{apa}

\usepackage{graphicx}
\usepackage{color}
\usepackage{amsmath}

\newcommand{\xmin}[0]{\ensuremath{x_f}}

\newcommand{\tstart}{\ensuremath{t_\textnormal{start}}}
\newcommand{\tend}{\ensuremath{t_\textnormal{end}}}
\newcommand{\tnow}{\ensuremath{t_\textnormal{now}}}

\newcommand{\alphaobs}{\ensuremath{\alpha_{\textnormal{obs}}}}%
\newcommand{\alphapred}{\ensuremath{\alpha_{\textnormal{pred}}}}%
\newcommand{\alphahat}{\ensuremath{\hat{\alpha}}}%
\newcommand{\alphaspre}{\ensuremath{\alpha(s; \deltapre)}}%
\newcommand{\alphaspost}{\ensuremath{\alpha(s; \deltapost)}}%
\newcommand{\alphas}{\ensuremath{\alpha(s; \Delta)}}%
\newcommand{\palpha}[1]{\ensuremath{p_\alpha\left(#1\right)}}

\newcommand{\ppre}[1]{\ensuremath{p_{i-1}\left(#1\right)}}
\newcommand{\ppost}[1]{\ensuremath{p_{i}\left(#1\right)}}

\newcommand{\deltapre}{\ensuremath{\Delta_{i}}}
\newcommand{\deltapost}{\ensuremath{\Delta_{i+1}}}

\newcommand{\xpre}{\ensuremath{{x_{i-1}}}}
\newcommand{\xzero}{\ensuremath{{x_i}}}
\newcommand{\xpost}{\ensuremath{{x_{i+1}}}}

\newcommand{\Deltamin}[0]{\ensuremath{\Delta_{\textnormal{min}}}}
\newcommand{\xmax}[0]{\ensuremath{x_{N}}}
\newcommand{\xcrit}[0]{\ensuremath{\hat{x}_s}}

\begin{document}

\title{Neural scaling laws for an uncertain world}
\author{Marc W.~Howard and Karthik H.~Shankar}
\affiliation{Department of Psychological and Brain Sciences, Center for Memory
and Brain, Initiative for Physics and Mathematics of Neural Systems, Boston
University}

\abstract{
Autonomous neural systems must efficiently process information in a wide range
of novel environments, which may have very different statistical properties.
We consider the problem of how to optimally distribute receptors along a
one-dimensional continuum consistent with the following design principles.
First, neural representations of the world should obey a neural uncertainty
principle---making as few assumptions as possible about the statistical
structure of the world.  Second, neural representations should convey, as much
as possible, equivalent information about environments with different
statistics.  The results of these arguments resemble the structure of the
visual system and provide a natural explanation of the behavioral
Weber-Fechner law, a foundational result in psychology.  Because the
derivation is extremely general, this suggests that similar scaling
relationships should be observed not only in sensory continua, but also in
neural representations of ``cognitive' one-dimensional quantities such
as time or numerosity.
}


\acknowledgements{
	We acknowledge helpful discussions with Eric Schwartz, Haim
	Sompolinsky, Kamal Sen, Xuexin Wei,  and Michele Rucci.
	This work was supported by BU's Initiative for the Physics and Mathematics
	of Neural Systems, NSF PHY 1444389, CRCNS NIMH R01MH112169, and NIBIB
	R01EB022864.	
}
\maketitle{}

The adapatability of mammals, humans in particular, to many novel environments
has been one of the keys to our evolutionary success.  
This flexibility is made possible, among other things, by the ability of the
sensory systems to represent information efficiently in a wide range of
circumstances.  For instance, consider the differing demands on the visual
system in a desert, or a dense forest, or an office environment
\cite{TorrOliv03}.  
One evolutionary strategy would be to adapt the properties of the sensors to a
particular environment, or a particular stimulus.  The ``fly detectors''
in the frog's brain described by Maturana and colleagues provides an early
example of this reasoning \cite{LettEtal59,MatuEtal60}.
However, if the efficiency of the visual system depended critically on the
statistics of a particular visual environment, it would leave the organism at
a disadvantage if it moved to a different environment, or if the environment
changed.
Of course, these considerations do not only apply to vision; analogous
arguments can be made for other sensory modalities and even neural
representation of more abstract quantities, such as time or numerosity.  This
suggests a design principle for the brain: the structure of neural
representations should make as few assumptions as possible about the
statistical structure of the world, which one might call a neural uncertainty
principle, and convey, as much as possible equivalent information about
environments with different statistics, which one might call a neural
equanimity principle.

We will argue in this paper that the Weber-Fechner law \cite{Fech60}, the
observation that the change in the perceptual magnitude of many
one-dimensional quantities\footnote{ The Weber-Fechner law does not apply to
circular variables, such as head direction or orientation.  }
depends on  the relative, rather than absolute change in the physical
quantity, can be understood as a behavioral manifestation of these design
principles.
The psychological scales of physical dimensions $x$ obeying the Weber-Fechner
law are logarithmic.  In the brain some external scalar quantities, such as weight
and luminance, may be expressed simply as the intensity of the response of a
single receptor.  This paper is concerned with variables, such as extrafoveal
retinal position, that are expressed by a distributed pattern of firing across
a set of receptors (which \citeNP{Stev57} referred to as metathetic
variables).  Each receptor is associated with a receptive field, which
describes the response of  the receptor to a constant stimulus with a
particular value $x$.  Each receptor responds to stimuli in a circumscribed
range of values of $x$; the array of all receptors can thus represent
functions over $x$, $f(x)$ \cite{MaEtal06}.  In this paper we study how to
optimally distribute receptors in order to represent arbitrary unknown
functions in the world.  Before formalizing this problem, we make some simple
mathematical observations about logarithmic scaling of receptors and then
briefly review the evidence that the visual system does (and does not) utilize
this form of neural scaling.

\begin{figure*}
	\begin{tabular}{lclc}
	\textbf{a} && \textbf{b}\\
		&
		\includegraphics[height=0.2\textheight]{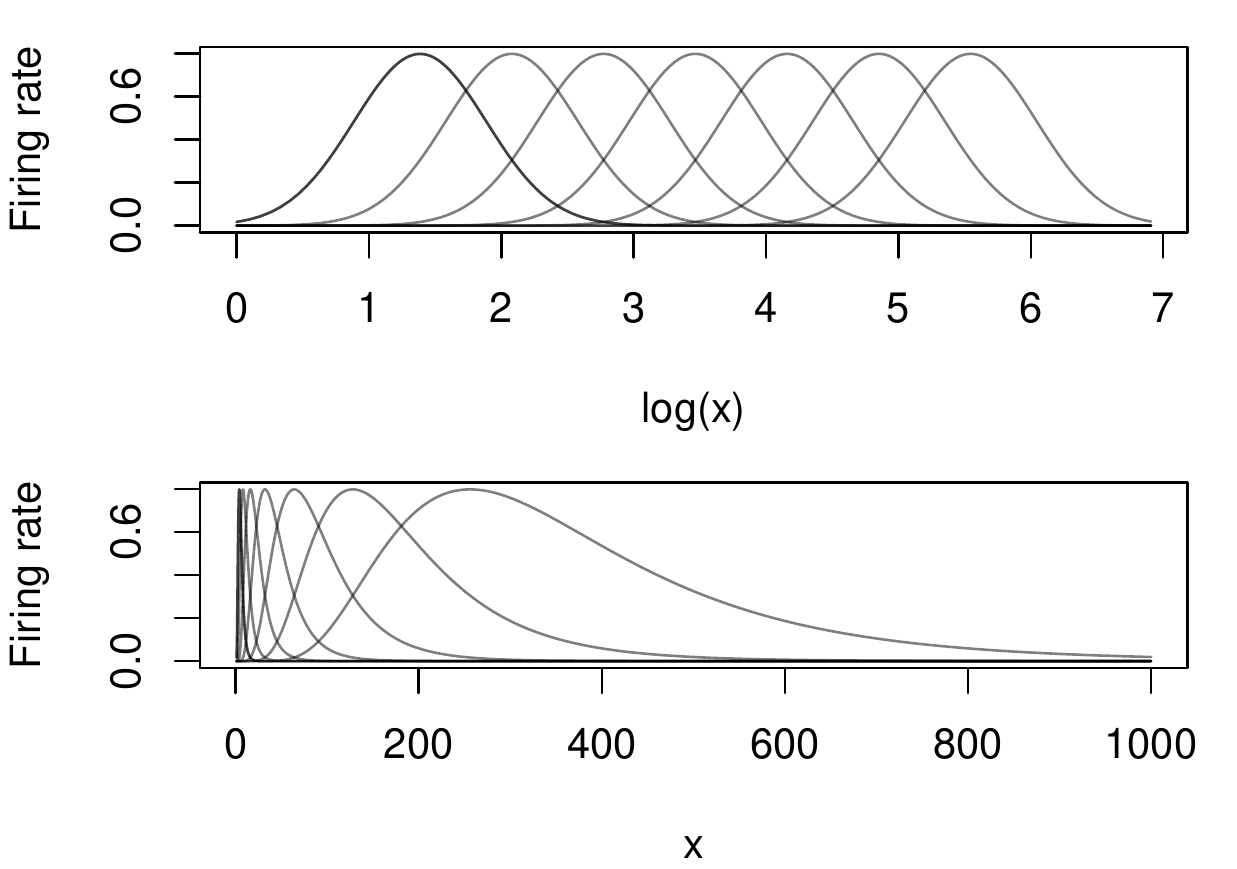}
		&& 
		\includegraphics[height=0.2\textheight]{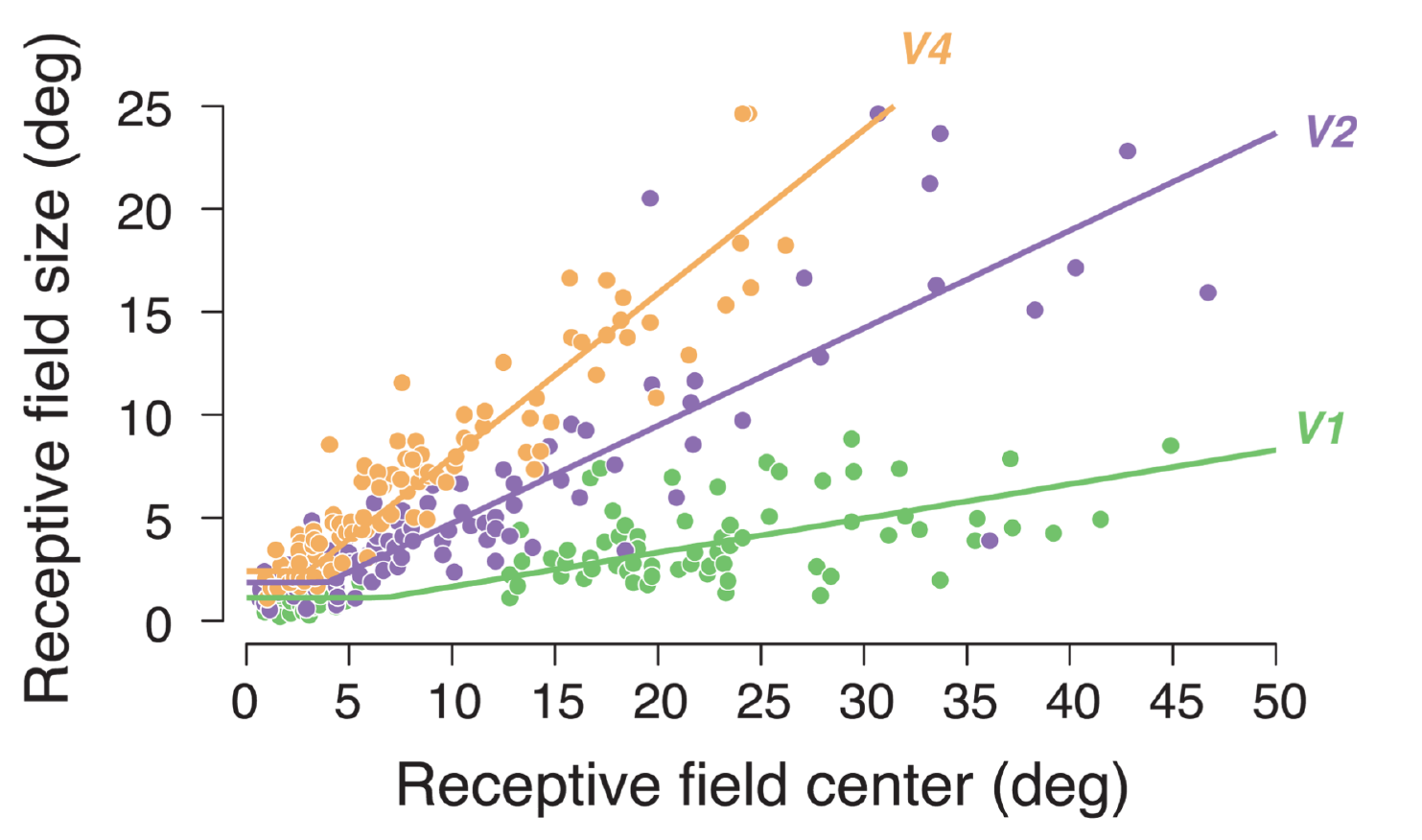}
	\end{tabular}
	\caption{
			\textbf{a.}  Weber-Fechner law can be generated from receptive
					fields that are of equal width and density as a function
					of $\log x$.  This results in receptive fields that
					increase in width and decrease in density as a
					function of $x$.  The increase in receptive field width
					should go up linearly with $x$; the density should go down
					like $1/x$.  The spacing of adjacent receptors should be a
					constant ratio for all values of $x$.
			\textbf{b.}
				Width of visual receptive fields as a function of eccentricity
						in three regions of macaque visual cortex.  
					The ``fovea'' can be seen as a region of constant spacing
					(flat line) near an eccentricity of zero.	
						There is a linear relationship outside of the fovea
						for all three regions.  Figure taken from Freeman \&
								Simoncelli (2011) who performed a metaanalysis
										on empirical studies cited
						therein.
			\label{fig:FreeSimo}
	}
\end{figure*}

\subsection{Logarithmic neural scales}
Receptive fields that are evenly-spaced and of equal width on a
logarithmic scale (Fig.~\ref{fig:FreeSimo}a, top) lead naturally to the
Weber-Fechner perceptual law.
A logarithmic scale implies several properties of the receptive fields
(Fig.~\ref{fig:FreeSimo}a bottom).  A logarithmic scale
implies that there should be fewer
receptive fields centered at high values of $x$, and 
receptors coding for higher values of $x$ should have wider receptive fields.    
These qualitative impressions can be refined to several quantitatively precise
relationships.  If the receptive fields are evenly-spaced  on a logarithmic
axis
then the width of receptive fields centered on a value $x$ should go up
proportional to $x$.  More generally, if the $i$th receptor has a receptive
field centered on $x_i$,
then the shape of the receptive
field should be constant for all receptors, but the width of the receptive
field should scale with the value of $x_i$.
If we denote the center of the $i$th cells receptive field as $x_i$ and define
$\Delta_i \equiv x_i - x_{i-1}$, then a logarithmic scale
implies that the ratio of adjacent receptors, $\Delta_{i+i}/\Delta_i$, is a
constant for all values of $i$.\footnote{Note that while logarithmic
scaling implies a constant ratio of adjacent receptor spacings, the
				converse is not true.  For
instance, if the ratio of receptor spacings was one, this would not lead to
logarithmic scaling.  Similarly, if we took a set of receptors on a
logarithmic scale and added a constant to each one, the ratio
$\Delta_{i+1}/\Delta_i$ would be unchanged and remain constant with respect to
$i$ whereas $x_{i+1}/x_i$ would no longer  be constant with respect to $i$.}
Finally we note that because $\log 0$ is
not finite, a logarithmic scale with a finite number of receptors cannot
continue to $x=0$.  In practice, this means that if the brain makes use of
logarithmic scales, there must be some other consideration for values near
zero.  In fact, as we will see below, the brain appears to solve this problem
in the visual system by including a region of approximately constant receptor
spacing for small $x$ (note the flat region around zero in
Fig.~\ref{fig:FreeSimo}b).

\subsection{Logarithmic neural scales in the mammalian brain}
There is strong evidence that the visual system in mammals obeys
logarithmic scaling in representing \emph{extrafoveal} retinal position.  In
monkeys, both the spacing of receptive fields (the cortical magnification
factor) \cite{DaniWhit61,HubeWies74,VanEEtal84,GattEtal88} and the
width of receptive fields \cite{HubeWies74,GattEtal88,Schw77} obey the
quantitative relationships specified by logarithmic scaling to a good degree
of approximation outside of the fovea.   
Figure~\ref{fig:FreeSimo}b (reproduced from \citeNP{FreeSimo11})  shows receptive
field width in the macaque across several cortical regions.  The fovea is
visible as a region of constant receptive field width, followed by a much
larger region showing a linear relationship between field width and
eccentricity, consistent with logarithmic scaling.  If the curves obeyed
logarithmic scaling precisely, the lines would continue to the origin rather than
flattening out in the fovea. However, because the spacing between receptors
would go to zero, this would require an infinite number of receptors.  

In the case of vision, logarithmic scaling can ultimately be attributed to
the distributions of receptors along the surface of the retina.  
However, this logarithmic mapping may be more general, extending to other
sensory \cite{MerzEtal73} and motor maps \cite{Schw77}.  In
addition, evidence suggests that logarithmic scaling holds for variables that
are not associated with any sensory organ.
Neural evidence from monkeys and humans
\cite{NiedMill03,HarvEtal13} suggests that neural representations
of non-verbal  numerosity are arranged on a logarithmic scale, consistent with
a broad range of behavioral studies of non-verbal mathematical cognition
\cite{GallGelm92,FeigEtal04}.  Behavioral and theoretical work
\cite{BalsGall09,HowaEtal15} suggests that a similar form of logarithmic
scaling could also apply to representations of time, which is
qualitatively consistent with growing body of neurophysiological evidence from
``time cells'' in the hippocampus, entorhinal cortex, and striatum
\cite{SalzEtal16,KrauEtal15,MellEtal15}.  
The neurophysiological evidence that these ``cognitive'' variables use
logarithmic scaling is not nearly as well-quantified as  visual receptive
fields.

\subsection{Overview}

If a set
of receptors was optimized under the expectation of a particular form of
statistics in the world, it would be suboptimal, perhaps disastrous, if the
organism encountered a state of the world with very different statistics.
The next section (``Optimal receptor distribution\ldots'') considers the problem of optimally placing receptors
to represent an unknown arbitrary function from a local perspective.  The
result of this section is that for all receptors to convey the same amount of
information as their neighbors, receptor spacings should be in a constant
ratio.  The following section~(``Global function representation'') considers
the amount of information conveyed by a set of receptors about a function
controlled by some fixed, but unknown, scale.  The global organization that
leads to equivalent information for a
wide range of scales has a region of constant spacing (ratio~1) followed by
a region with constant ratio spacing (ratio~$>1$) which resembles the
organization of the visual system with a closely-packed fovea surrounded by
a logarithmic scale.
  
\section{Optimal receptor distribution for representing arbitrary unknown functions}
\label{sec:local}

We want to represent a function in the world over some continuous real value
$x$.  Receptors sample the function in some neighborhood such that each
receptor is ``centered'' on a particular value of $x$.
The goal is to distribute the receptors to enable them to represent an
arbitrary and  unknown function $f(x)$.   
In order for this problem to be meaningful we must assume that there is a
finite number of receptors.  We assume here that the lower and upper bounds of the
scale $x$ are physically constrained.  At the lower bound, perhaps there is
some minimal degree of resolution that can be achieved with the receptors.  In
some cases the upper bound may be given by the properties of the world or
anatomy.  For instance, in vision, ecccentricity cannot possibly be larger
than $\pi$.  Our question, then is how to distribute the location of the
receptors within the range of $x$ values to be represented.

Each additional receptor provides an additional benefit to the organism to the
extent that it provides additional information  about the function.
Intuitively, if we clumped all of the receptors close together this would be
suboptimal, because the receptors would all be conveying the same information
and there would be no receptors to communicate information about other regions
of the function.  Naively, we might expect that the solution is
trivial---perhaps receptors should be evenly-spaced along the $x$-axis.  It
will turn out that this is not in general the optimal solution.  Rather, the
spacing of each pair of receptors should be in a constant ratio to the spacing
of the previous pair of receptors. Constant spacing corresponds to
a ratio of~1; it will turn out that other ratios are admissable as well.

In order to do the calculation we need to measure the redundancy between a
pair of receptors.   There are a number of measures of redundancy one might
use.  For instance, we might measure the mutual information between the output
of two receptors after observing the world for some period of time.   There are
other measures that one might use.  All things equal, we would expect the
redundancy between two receptors to be higher for receptors that are placed
close together rather than receptors that are placed far apart.  The
development below applies to any measure of redundancy that obeys some basic
properties.\footnote{After the general derivation we include a worked example
with a particularly tractable measure of redundancy.}
The distribution of receptors is optimal if each receptor is expected to be as
redundant with its predecessor as it is with its successor.  If that was not
the case, then we could move one of the receptors and get more non-redundant
information out about the function.  

\subsection{Formulation of the problem}

\begin{figure}
	\centering
	\begin{tabular}{lc}
	\textbf{a}\\
   & 	\includegraphics[width=0.3\columnwidth]{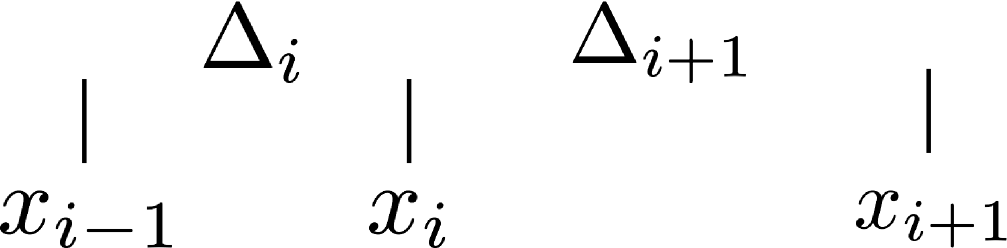}\\
   ~\\
	\textbf{b}&\\
   & 	\includegraphics[width=0.55\columnwidth]{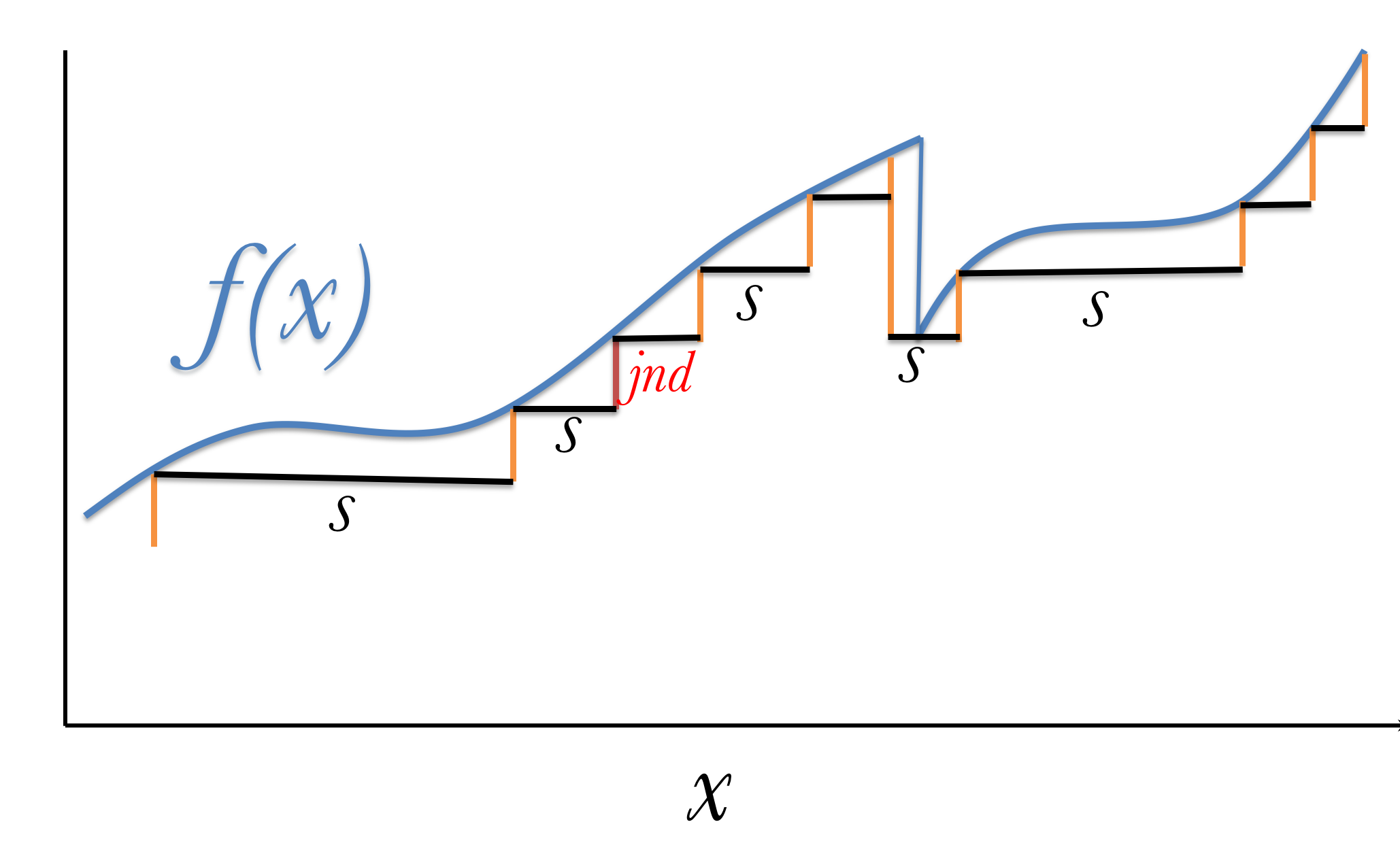}
   \end{tabular}
		\caption{
		\textbf{a.} Discretizing the function.  After picking some small
				resolution (``jnd'' in the figure), we discretize the signal.
				This defines a variable $s(x)$ describing the distance (or
								scale) between $x$ and the next location where
				the function takes on a different discretized value.
		\textbf{b.} Schematic for notation.
		The derivation assumes that there are receptors at $\xpre$ and
				$\xzero$ separated by $\deltapre$.  The goal is to choose
				$\deltapost$ (and thus $\xpost$) such that the redundancy
				between the pairs of receptors is equated.
				\label{fig:xzeroschematic}
	}
\end{figure}

Suppose we have set of receptors with receptive fields centered on positions
$x_1$, $x_2$, \ldots $x_N$.   As before, let us denote the distance between the
locations of receptor $\xpre$ and $\xzero$ as $\deltapre$.   Constant receptor
spacing would imply that $\deltapre = \deltapost$ for all $i$.  It will turn
out that this is not the general solution to our problem.  
Rather, the solution is to space the receptors such that the ratio of
adjacent receptor spacings is a constant across the scale.  That is,
the ratio $\deltapost/\deltapre$ takes the same value for all $i$.

We assume that in
the neighborhood of each $x_i$, the function is controlled by some scale
$s_\xzero$.   To quantify this in an unambiguous way, assume that our
receptors can only respond over a finite range with some non-zero resolution. 
Accordingly, we discretize the function to be represented to some degree of
resolution (Figure~\ref{fig:xzeroschematic}a) so that at each value of $x$,
$s_x$ gives the distance to the next value of $x$ where the discretized
function would take a different value than the discretized $f(x)$.  That is,
for each $x$, the discretized function $f(x) = f(x')$ for each $x < x' < x +
s_x$.  

Note that if we knew the scale $s_\xzero$ at $\xzero$, it would be
straightforward to estimate the redundancy between the receptor at $\xzero$ and
the receptor at $\xpost$. If we knew that the spacing between receptors
$\deltapost$ were much smaller than $s_\xzero$, the receptors would measure
the same value and we would expect the redundancy to be high.  In contrast,
if we knew that the spacing between the receptors were much
smaller than $s_\xzero$ then the value of the function would have changed at
least once between $\xzero$ and $\xpost$ and we would expect the redundancy to
be low.  Let us denote the function relating the redundancy we would observe
between receptors spaced by a particular value  $\deltapre$ if we knew the
scale took a particular value $s$ as $\alphaspre$.  We assume $\alphaspre$
to be a monotonically decreasing function of its argument.  Note that we
could choose any number of different ways to measure redundancy.  For
instance, we could treat mutual information as our measure of redundancy and
this, coupled with knowledge of the properties of the receptors, would specify
a specific form for $\alphaspre$.  But we could just as well measure
redundancy in other ways or assume different properties of our receptors,
which would result in a different form for $\alphaspre$.  The arguments below
about optimal receptor spacing apply to any measure of redundancy that obeys
some basic properties described below and does not depend critically on how
one chooses to measure redundancy. 

Observing the redundancy between the receptors at $\xpre$ and $\xzero$ 
spaced by  $\deltapre$ allows us to infer the value of the scale at the first
receptor $s_\xpre$.  For instance, if the redundancy is maximal, that implies
that the function did not change between $\xpre$ and $\xzero$.  In contrast,
if the redundancy between the receptors at $\xpre$ and $\xzero$ is very small,
that implies that the scale of the function at $\xpre$, $s_\xpre$ was smaller
than $\deltapre$.  Knowing the scale at $\xpre$ places constraints on the
value at $\xzero$.  For instance, if we knew with certainty that $s_\xpre =
\deltapre + C$, then we would know with certainty that $s_\xzero = C$.  So, if
we knew the probability of each value of $s_\xpre$ this constrains the
probability distribution for $s_\xzero$.  

Let us fix the spacing $\deltapre$ between the receptors at $\xpre$ and
$\xzero$, sample the world for some time (observing many values of the
the receptor outputs in response to many samples from the function)
and denote the average value of redundancy we observe as $\alphaobs$.  Denoting
the (unknown) probability of observing each possible value of $s_\xpre$ as
$\ppre{s}$, then the observed value of redundancy, $\alphaobs$, resulted from
an integral:
\begin{equation}
		\alphaobs = \int \ppre{s}\ \alphaspre\ ds
			\label{eq:alphaobs}
\end{equation}
This can be understood as an inference problem; observation of $\alphaobs$
leads us to some belief about the distribution $\ppre{s}$.    Informed by this
knowledge, we then place the third receptor at a spacing $\deltapost$.  If we
knew the distribution of scales at $\xzero$,  $\ppost{s}$, then we would
expect to observe  a redundancy of 
\begin{equation}
		\alphapred = \int \ppost{s}\ \alphaspost\ ds
			\label{eq:alphapred}
\end{equation}
between the second pair of receptors.  Our problem is to choose $\deltapost$
such that we would expect $\alphapred = \alphaobs$.  

\subsection{Minimal assumptions about the statistical properties of the world}
The actual value of $\deltapost$ that makes $\alphapred=\alphaobs$ of course
depends on the detailed properties of the receptors, the function
$\alphas$ and the statistics of the world.  We show, however, that for
minimal assumptions about the world, this problem results in the solution
that, whatever the value of $\alphaobs$ one finds that for some spacing
$\deltapre$, the choice of $\deltapost$ is such that only
$\deltapost/\deltapre$ is affected by the observed value $\alphaobs$.

First, we make minimal assumptions about the function to be estimated.
We assume an uninformative prior for $\ppre{s}$.
We make the minimal assumptions that successive values of $s$
are independent, and that the value of the function past the $s_x$ is
independent of the value at $x$.    That is we assume that $f(x)$ and
$f(x+s_x)$ are independent of one another, as are  $s_x$ and $s_{x+s_x}$.
\footnote{The
critical point of these assumptions is that at no point to we introduce an
assumption about the statistics of the world that would fix a scale for our
receptors.  For instance, the result of receptor spacings in a constant ratio
would also hold if we assumed a  power law prior for $\ppre{s}$ rather than a
uniform prior because the power law is a scale-free distribution.  Similarly,
it is acceptable to relax the assumption of independence between $f(x)$ and
$f(x+s_x)$ as long as doing so does not introduce a scale.
\label{foot:power}
}

Second, we require that the receptors do not introduce a scale \emph{via}
$\alphas$.  In general, the function $\alphas$ will depend on how we choose to
quantify redundancy   and the properties of the receptors.  However, as long
as $\alphas$ rescales, such that it can be rewritten  in a canonical form
\begin{equation}
	\alphas = \alphahat(s/\Delta),
\label{eq:alphahat}
\end{equation}
the conclusions in this section will hold.  One can readily imagine idealized
settings where Eq.~\ref{eq:alphahat} will hold.   For instance, we could
assume we have perfect receptors that sample the function at only one point
and take our measure of redundancy to be one if the receptors observe  the
same value up to the resolution used to specify $s$
(Fig.~\ref{fig:xzeroschematic}b) and zero otherwise.  More generally, for
imperfect receptors that sample a range of values, Eq.~\ref{eq:alphahat}
requires that the receptive fields scale up with $\Delta$.  This assumption is
necessary, but not sufficient to equate redundancy between receptors.

Our third requirement is referred to as the Copernican principle. Because it is
critical to the argument in this section, we discuss it in some detail in the
next subsection.

\subsection{The Copernican Principle}

In addition to minimal  assumptions about the statistics of the world and the
properties of the receptors, we assume that the world's choice of $s_\xpre$ is
unaffected by our choice of $\deltapre$. The belief that there is nothing
privileged about one's point of observation is referred to as the Copernican
principle. The Copernican principle  was employed by \citeA{Gott93} to
estimate a probability distribution for the duration of human civilization.
Because this provides a concrete illustration of an important concept, it is
worth explaining Gott's logic in some detail.

Suppose we observe Fenway Park in 2017  and learn it has been standing for 103
years.  Knowing nothing about construction techniques or the economics of
baseball we want to estimate how much longer Fenway Park will stand.   Because
there is nothing special about our current viewpoint in 2017, our observation
of Fenway should be uniformly distributed across its lifetime.  That is, the
observation at \tnow{} in 2017 ought to be uniformly distributed between
$\tstart$ when Fenway Park was constructed (in 1914) and the unknown time of
its destruction $\tend$.
Gott argued that we should expect $\tnow$ to be uniformly distributed between
$\tstart$ and $\tend$, such that we should expect  $\tend - \tnow$ to be longer than
$\tnow-\tstart$ half the time.     This means that the age
of the structure (102 years) fixes the units of the distribution of $\tend -
\tstart$.  If we observed an object that has been existing for 1030 years, or
for 103 seconds, then our inference about its expected duration would have the
same shape, but only differ in the choice of units.  

The Copernican argument also applies to our inference problem.  Here the scale
at the first receptor $s_\xpre$  plays a role analogous to the duration of
the object $\tend - \tstart$. 
Suppose that we have a set of receptors with a known function $\alphas$  that
obeys Eq.~\ref{eq:alphahat} and we encounter a world with some unknown
statistics.  In this first world we choose some $\deltapre$, observe some
measure of redundancy $\alphaobs$, use some method of inference to estimate
$\ppre{s}$ and $\ppost{s}$ and then select the value of $\deltapost$ to yield the
expectation that $\alphapred=\alphaobs$.  Let us refer to the distribution of
$\ppost{s}$ that we inferred from fixing $\deltapre$ to its particular value and
observing $\alphaobs$ as $\palpha{s;\deltapre}$.  Now, suppose that we
encounter another world with the same receptors.  Assume further that we
choose a different $\deltapre$ but observe the \emph{same} value of
$\alphaobs$.  How should our inference about the distribution of $s_\xzero$ be
related to our inference from the first world?  If Eq.~\ref{eq:alphahat}
holds, the Copernican Principle requires that all of the distributions that can
be inferred for a particular value of $\alphaobs$ must be related to one
another \emph{via} a canonical form: 
\begin{equation} 
		\palpha{s;\deltapre} = \frac{1}{\deltapre} \hat{p}_\alpha(s/\deltapre).  \label{eq:pcanon}
\end{equation} 
If Equation~\ref{eq:pcanon} did not hold, it would imply that we can infer
something about the world's choice of $s_\xzero$ from our choice of
$\deltapost$ \emph{beyond} that communicated by the value of $\alphaobs$.
As long as the receptors and our measure of redundancy scale with $\deltapre$ as
in Eq.~\ref{eq:alphahat} then any such effect would imply that the world's
choice of $s$ and our choice of $\deltapre$ are not independent and thus
violate the  Copernican Principle.  

\subsection{Ratio scaling results from minimal assumptions about the
statistics of the world and the Copernican Principle}

Equations~\ref{eq:alphahat}~and~\ref{eq:pcanon} imply that redundancy is
equated for receptor spacings in some ratio.   To see this, let us rewrite
Eq.~\ref{eq:alphapred} and find
\begin{eqnarray}
	\alphapred(\deltapost) &  = & 
			\frac{1}{\deltapre}
			\int_0^\infty 
				\hat{p}_\alpha\left(s/\deltapre\right) \ 
						\alphahat(\deltapost/s)  \
			ds \nonumber\\
			& = & 
			\int_0^\infty 
				\hat{p}_\alpha(s') \ 
					\alphahat(r/s') \
			ds'
			\label{eq:canon}
\end{eqnarray}
In the second line $s' \equiv s/\deltapre$ and 
we define the ratio $r \equiv \deltapost/\deltapre$.  
The right hand side of Equation~\ref{eq:canon} clearly does not depend on
$\deltapost$ directly, but only on the ratio.   

The finding, then, is that a principle of minimal assumptions about the
statistics of the world coupled with the Copernican Principle implies that the
optimal distribution of receptor locations is such that the ratio of
successive receptor spacings is constant.  Logarithmic neural
scales also imply that the ratio of successive receptor spacings are constant.
In this sense, logarithmic scales can be understood as a response to the
demand that the receptor layout are expected to equalize redundancy across
receptors in a world with unknown statistics.  Because logarithmic scales do
not uniquely predict constant ratio spacing this cannot be the entire story.
We pursue additional constraints that imply logarithmic scales, with some
important exceptions when $x$ approaches zero, in the next section.  Before
that, for concreteness we include a worked example for a idealized set of
receptors and a specific choice for $\alphas$.

\subsection{A worked example}

The general development above provides a set of conditions that result in
a constant ratio of spacing between receptors.  In order to make this more
concrete, we work out a specific example with specific choices for the
properties of the receptors, the measure of redundancy and the method of
inference.  

This example assumes that the receptors are perfect and the $i$th receptor
samples the function only at the location $x_i$.  For simplicity we define a
measure of redundancy that gives $\alpha = 1$ if the two receptors observe the
same value of $f$ and $\alpha=0$ otherwise.  This lets us write out
\[
\alphaspre 	= 
			\begin{cases}  
					1, & \mbox{if} \ \deltapre < s_\xpre\\
					0, &   \mbox{if}\ \deltapre \ge s_\xpre
			\end{cases}
\]
The
same holds for $\alphaspost$, only comparing $\deltapost$ to $s_\xzero$.
With our simplified definition of redundancy, any value of $\alphaobs$ that is
not zero or one must have resulted from a mixture of those two cases with
probability $\alphaobs$ and $1-\alphaobs$ respectively.  Let us first consider
the case where $\deltapre < s_\xpre$, as it aligns perfectly with the Gott
argument.

If $s_\xpre > \deltapre$, then  $s_\xpre$ plays the role of the unknown
duration of the lifetime of an object such as Fenway Park, $\tend-\tstart$.
The difference between the two receptors $\deltapre$ plays the role of the
time of the current observation $\tnow - \tstart$ and the unknown $s_\xpost$
is analogous to $\tend - \tnow$.
Gott's calculation defines the ratio $r=\left(\tend -
\tnow\right)/\left(\tnow-\tstart\right)$.  The Copernican Principle leads to
the belief that
$\tnow$ ought to be uniformly distributed between $\tstart$ and $\tend$.
Thus the cumulative distribution of $r$ obeys:
\begin{equation}
P\left(r > Y\right) = \frac{1}{1+Y}
\label{eq:Gott}
\end{equation}
Note that the elapsed duration $\tnow - \tstart$ only enters this expression
\emph{via} the ratio.

Similar arguments apply to the spacing of receptors.   
If we observe $\alphaobs=1$, then the posterior $\ppost{s}$ 
is controlled by Eq.~\ref{eq:Gott}, with
$\deltapre$ in place of $\tnow - \tstart$ and $\deltapost$ in place of $\tend
- \tnow$.\footnote{That is, Eq.~\ref{eq:Gott}
		gives the cumulative of the posterior distribution.}   In this toy
		problem, the value of
		$\alphapred$ is given by Eq.~\ref{eq:Gott} if $\alphaobs=1$.
If instead of $\alphaobs =
1$,  we observed  $\alphaobs=0$, we would infer a uniform distribution of
$s(\xzero)$.\footnote{One can argue for other ways to make the inference in
this toy problem when $\alphaobs=0$.  As long as those depend
only on the ratio $\deltapost/\deltapre$ and not explictly on $\deltapost$,
the conditions of the general argument still hold.} If 
our prior on the distribution of $s_\xzero$ is uniform, then $\alphapred$ will
be one for any finite value of $\deltapost$.

After sampling the world for some period of time, if we observe $\alphaobs$
as a number between zero and one, then our posterior distribution of scales
should be a mixture of the inference from the two cases and simplify
Eq.~\ref{eq:alphapred} as: 
\begin{eqnarray}
	\alphapred(\deltapost) =
	\alphaobs  \frac{1}{1+\deltapost/\deltapre} + \left(1-\alphaobs\right)
	\label{eq:simpleequation}
\end{eqnarray}
It is clear that the right hand side is only a function of the ratio
$r \equiv \deltapost/\deltapre$ so that the value of $\deltapost$ that makes
$\alphapred=\alphaobs$ depends only on the ratio.\footnote{In
this simple example  there is a solution for values of $\alphaobs > 1/2$. }
For any value of $\alphaobs$, the same value of $r$ satisfies this equation
for any choice of $\deltapre$.  In this simple problem we see that the choice of
$\deltapre$ can only affect the answer by
fixing its units.  This naturally results in the
ratio of adjacent receptor spacings being constant, as implied by logarithmic
receptor scales.


\section{Global function representation} 
\label{sec:global}
\begin{figure}
	\begin{tabular}{lclc}
		\textbf{a} & & \textbf{b}\\
				&\includegraphics[width=0.4\columnwidth]{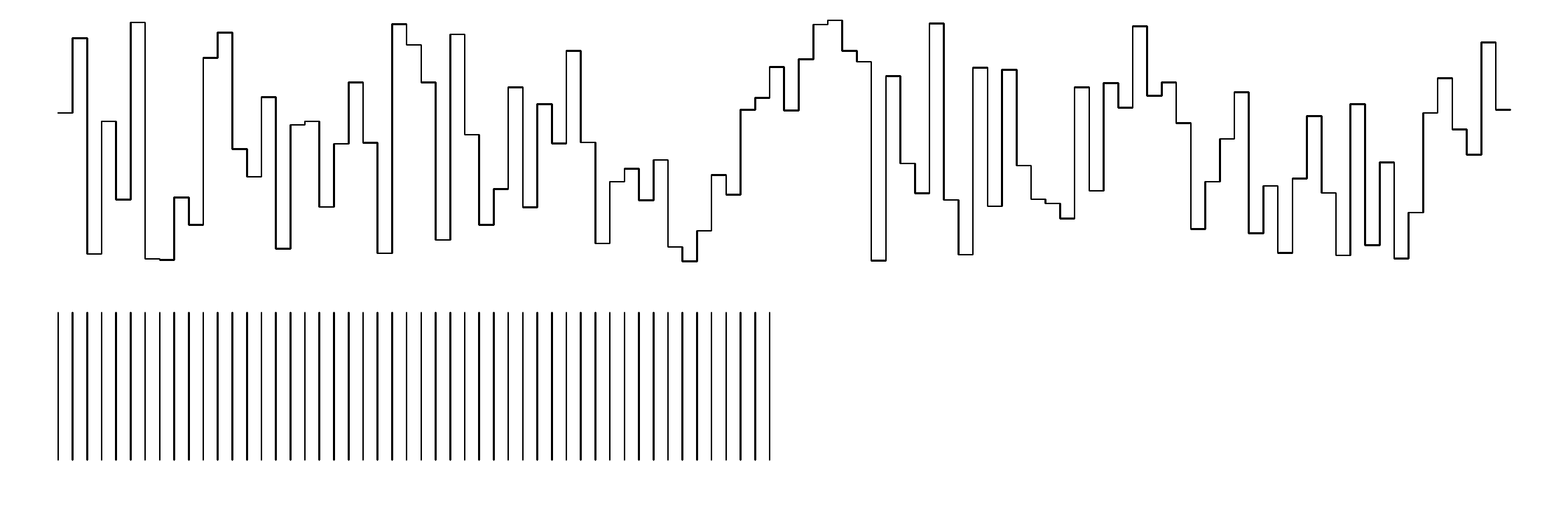}
				&&\includegraphics[width=0.4\columnwidth]{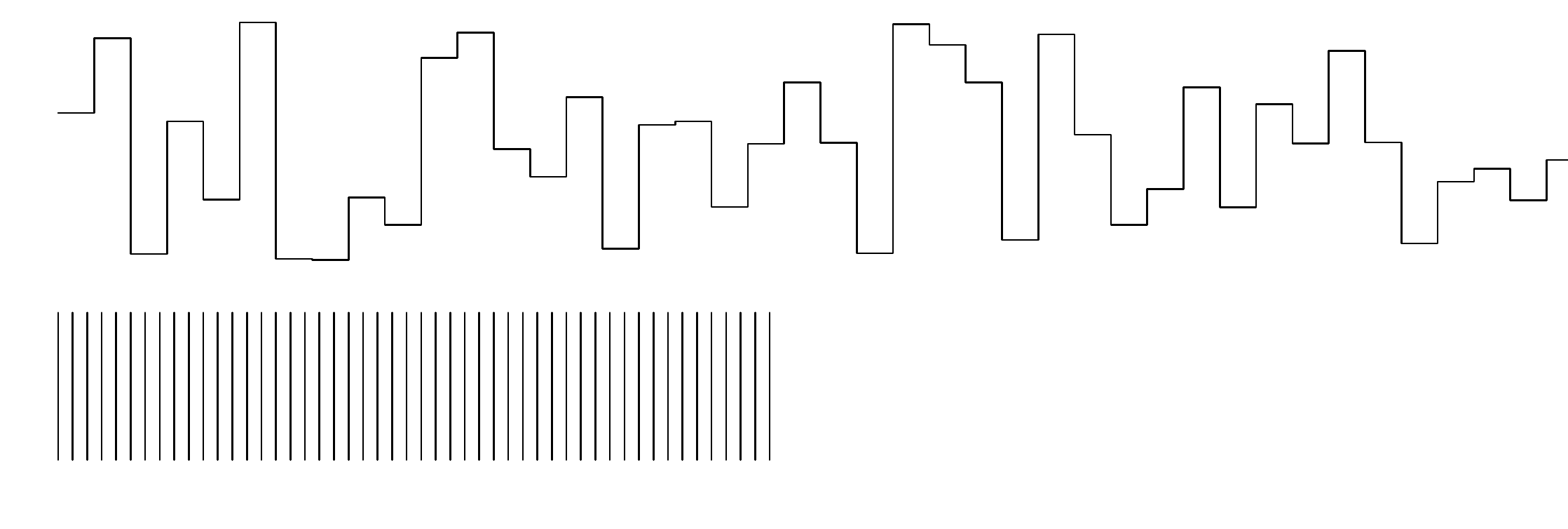}\\
		\textbf{c} & & \textbf{d}\\
				&\includegraphics[width=0.4\columnwidth]{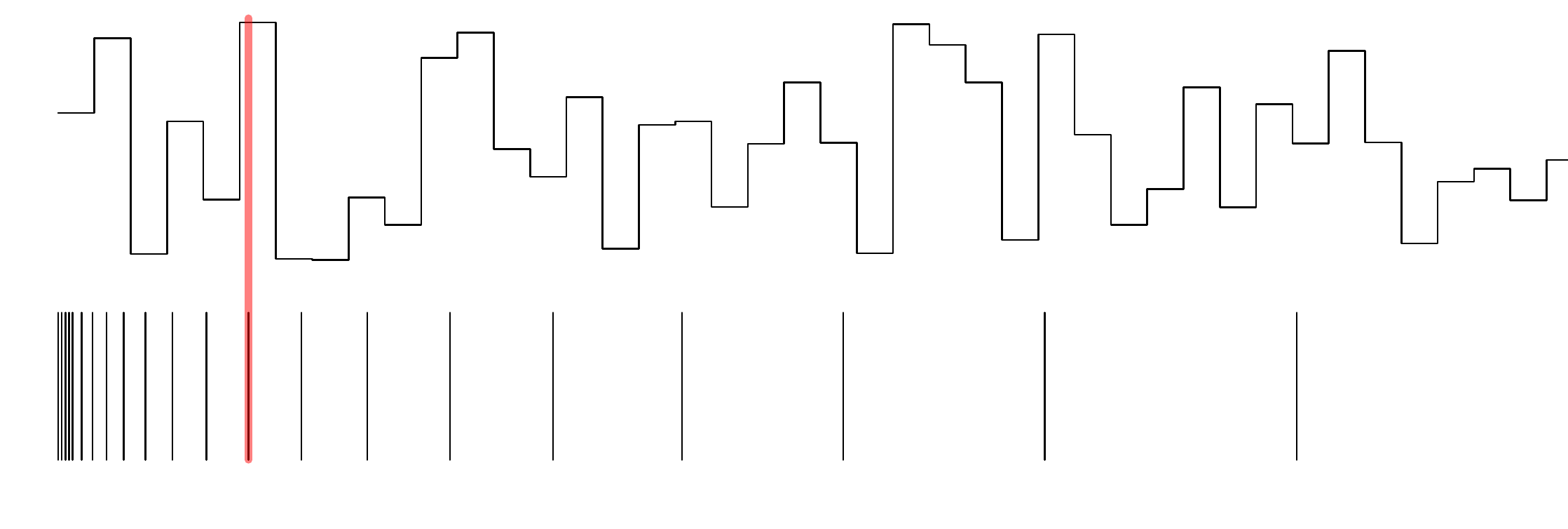}
				&&\includegraphics[width=0.4\columnwidth]{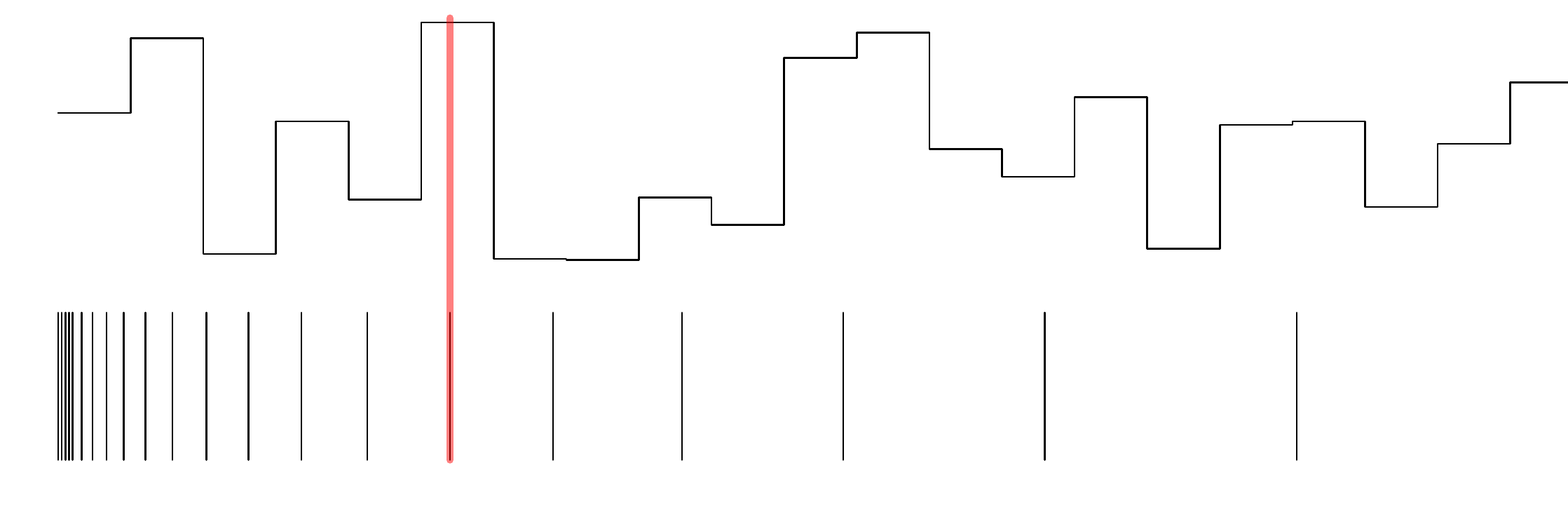}
	\end{tabular}
	\caption{ { \bf Cartoon illustrating the interaction of receptor spacing
			and function scale.}
	In each panel, the top curve shows a specific function $f_s(x)$; the
			functions in the four panels differ only in their scale $s$.  On
			the bottom of each panel, the vertical lines display receptor
			locations.  {\bf a-b.} Constant spacing.  When $c=0$,
		the receptors are evenly-spaced, fixing a scale to the receptors.  If
		the function has the same scale as the receptors, as in \textbf{a},
		each receptor captures a different value of the function, the
		veridical region runs the entire length of the receptor array
		and the information transmitted is maximal.  However, if the scale of
		the function is larger than the constant spacing, as in \textbf{b},
		the array of receptors conveys less information because there is less
				information in the function
		over that range.
		\textbf{c-d.}  Ratio spacing. When $c > 0$, the receptors can 
		extend over a much wider range of $x$ values than with constant
		spacing.  The receptors carry information about the function
		over a veridical region where $\Delta_i < s$; the border of the
		veridical regions is shown (in cartoon form) as a vertical red line.
		As $s$ increases (from \textbf{c} to \textbf{d}), the size of the
		veridical region in $x$ changes, but as long as the border is covered
		by the receptors, the amount of nonredundant information in the
		veridical region is constant as a function of $s$. This property does
		not hold if the scale of the function approaches the minimum receptor
		spacing.
\label{fig:adaptive}}
\end{figure}

The foregoing analysis conducted at the level of pairs of receptors showed
that the optimal spacing in order to represent arbitrary functions places the
receptors in constant ratio, but does not specify the value of the ratio.
It is  convenient to parameterize the ratio between adjacent receptors by a
parameter $c$ such that $r \equiv \frac{\deltapost}{\deltapre} = 1+c$.  If
$c=0$, receptor spacing is constant; if $c>0$ the ratio is constant, as in a
logarithmic scale.  We consider the
global coding properties of these two schemes as well as a hybrid scheme in
which the first part of the axis has constant spacing ($c=0$) followed by a
region with logarithmic spacing ($c>0$), analogous to the organization of the
visual system, with the region of  constant spacing corresponding roughly to
the fovea (Fig.~\ref{fig:FreeSimo}b).

\subsection{Formulating the problem}
For simplicity, we assume that each receptor
perfectly samples the value of a function in the world at a single
perfectly-specified location and consider a simple class of functions
$f_s(x)$, which consist of independently chosen values over the range 0~to~1
at a spacing of $s$ such that $f_s(x)$ and $f_s(x+s)$ are independent and
$f_s(x)$  and $f_s(x+s - \epsilon)$ are identical for $\epsilon < s$.  The
information about a function contained in a given range by a particular instantiation of the
function is just the number of entries specifying the values of the function
over that range.  Note that $s$ controls the density of information conveyed
by $f_s$ over a given range of $x$.  We assume that $f_s(x)$ is specified over
the entire range of $x$ from zero to infinity so that the total amount of
information that could be extracted from a function is infinite for every
finite scale.  Figure~\ref{fig:adaptive} shows several functions and choices
of receptor scaling in cartoon form. 

A set of receptors can do a ``good'' job in representing a function $f_s(x)$
as long as the spacing between the receptors is less than or equal to $s$.
With the simplifying assumptions used here, each of the non-redundant values
of the function is captured by at least one receptor allowing reconstruction
of the function with error in the $x$ location no worse than $s$.  More
generally, even with coarse-graining and noisy receptors it is clear that
there is a qualitative difference between the ability of the receptors to
measure the function when $\Delta < s$ compared to when $\Delta > s$.
Let us define the veridical
region of the function as the range of $x$ over which $\deltapre < s$.  We
also assume there is some minimum possible receptor spacing $\Deltamin$.  We
measure the amount of veridical information $I$ conveyed by the set of
receptors as the number of unique function values that the function has within
the veridical region.

\subsection{Constant receptor spacing, $c=0$}
\begin{figure}
	\centering
	\includegraphics[width=0.8\columnwidth]{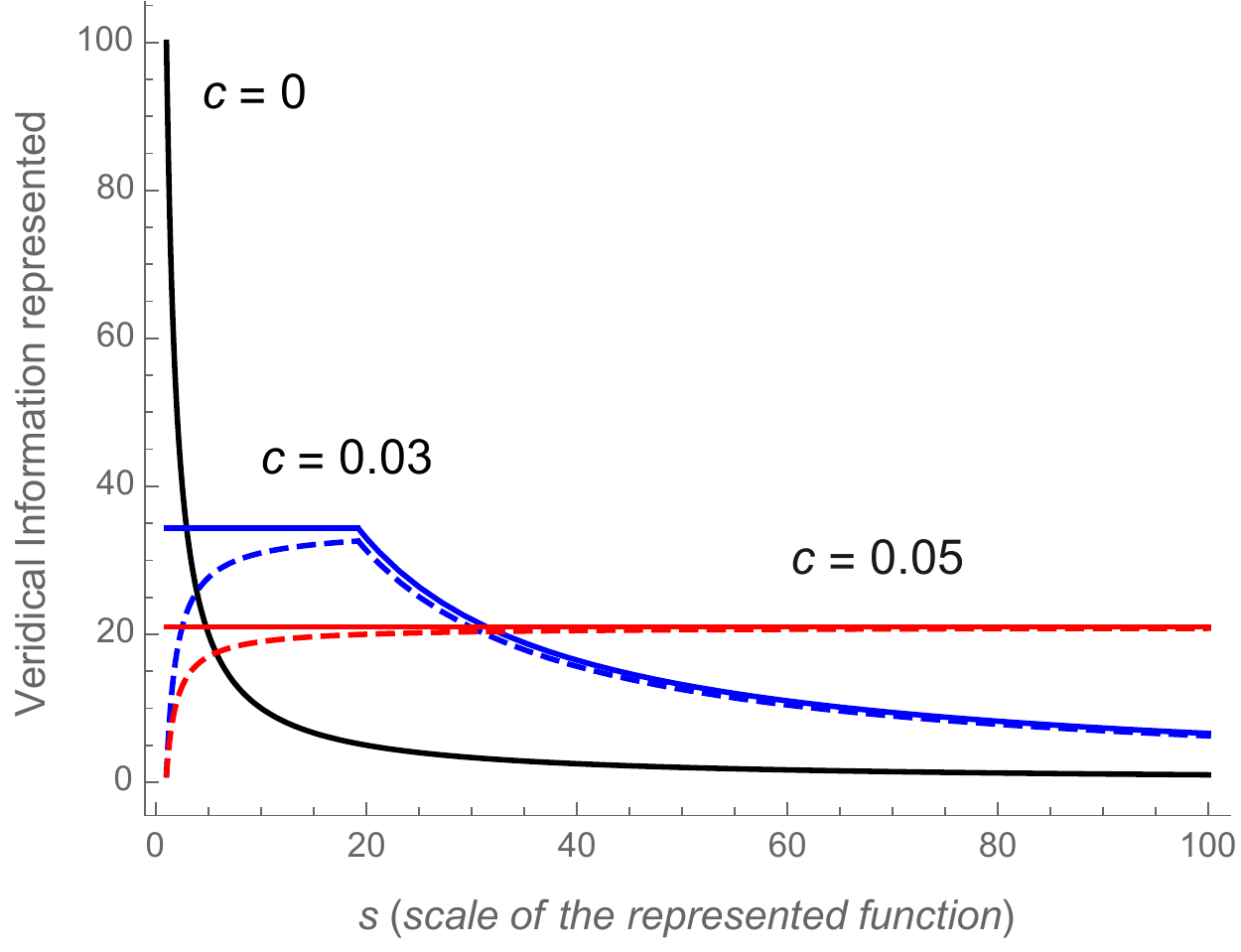}
	\caption{\textbf{Information in the veridical region as a function of
			scale for three forms of receptor spacing.}
			The curve labeled $c=0$ shows the results for constant spacing.  This
					falls off rapidly for scales larger than the scale set by
					the receptor spacing.  Solid curves show results for ratio
					spacing ($c>0$) with a fovea; dashed curves show results
					for ratio spacing without a fovea.  
					$N=100$ for all curves.  
							See Figure~\ref{fig:schematicfovea}a for an intuitive
							explanation of why the information decreases for
							small values of $s$.
							The fovea has 
					an additional $1/c$ receptors. 
					Note that there is a region where the
					amount of information conveyed is approximately constant
					as a function of scale before falling off as the veridical
					region exhausts the number of receptors $N$.  For $c=.05$,
					 this region is off the scale of this figure.  Note also
					 the error at small scales for $c>0$ without a fovea.
				\label{fig:informationc}
	}
\end{figure}

First, consider the implications of constant receptor spacing.
If we knew the value of $s$ controlling $f_s(x)$, placing our receptors with 
constant separation $\Delta=s$ would be sufficient to convey all of the
information in the function over the entire range of $x$ values covered by the
receptors. 
$N$ receptors would be able to accurately represent the
function over a veridical region of width $N\Delta $ and the amount of information
conveyed about the function in that region is just $N$.  However, our constant
choice of spacing would have poor consequences if  our choice of $\Delta$ did
not correspond to the world's choice of $s$.  
If 
$s > \Delta$, the veridical region is still of length $N \Delta$, but the
amount of information contained in that region is only $N\frac{\Delta}{s}$,
decreasing dramatically as $s$ increases (curve
labeled $c=0$ in Figure~\ref{fig:informationc}).  With constant spacing, $c=0$,
the amount of information conveyed by the receptors depends dramatically on
$s$.

\subsection{Constant receptor spacing ratio, $c>0$}
\begin{figure}
	\begin{center}
	\begin{tabular}{lclc}
		\textbf{a} && \textbf{b}\\
			&\includegraphics[width=0.4\columnwidth]{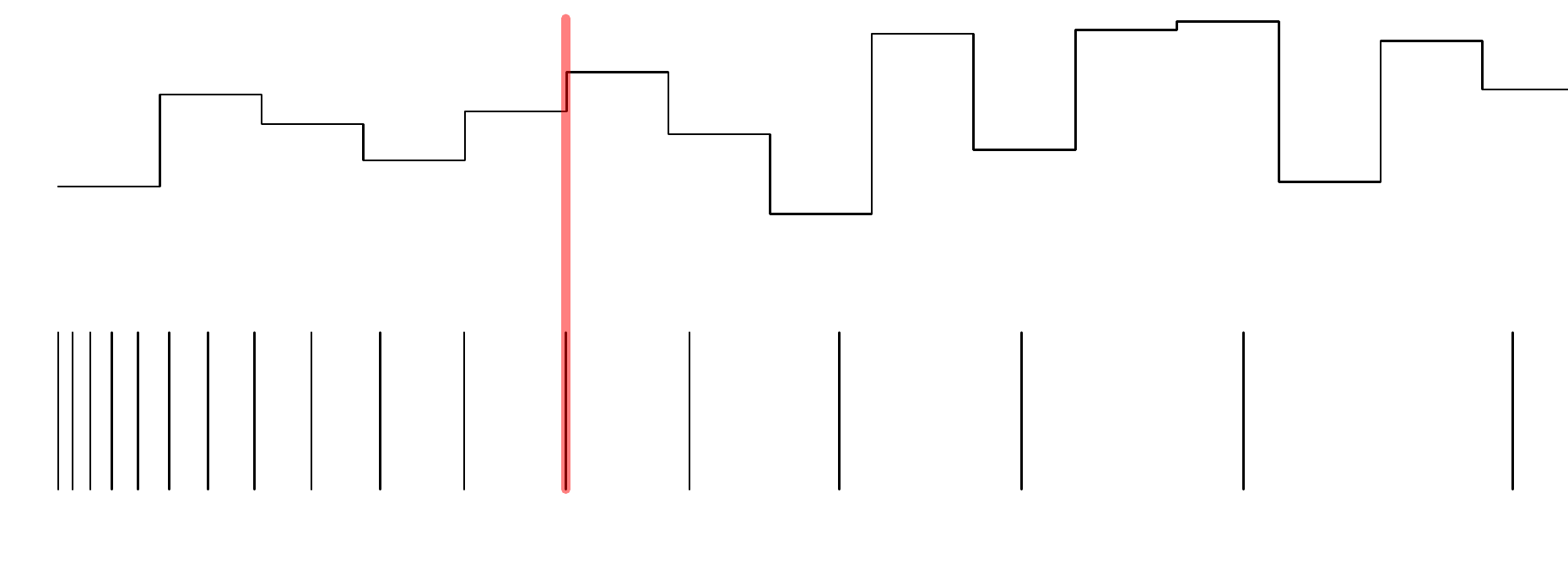}
			&&\includegraphics[width=0.4\columnwidth]{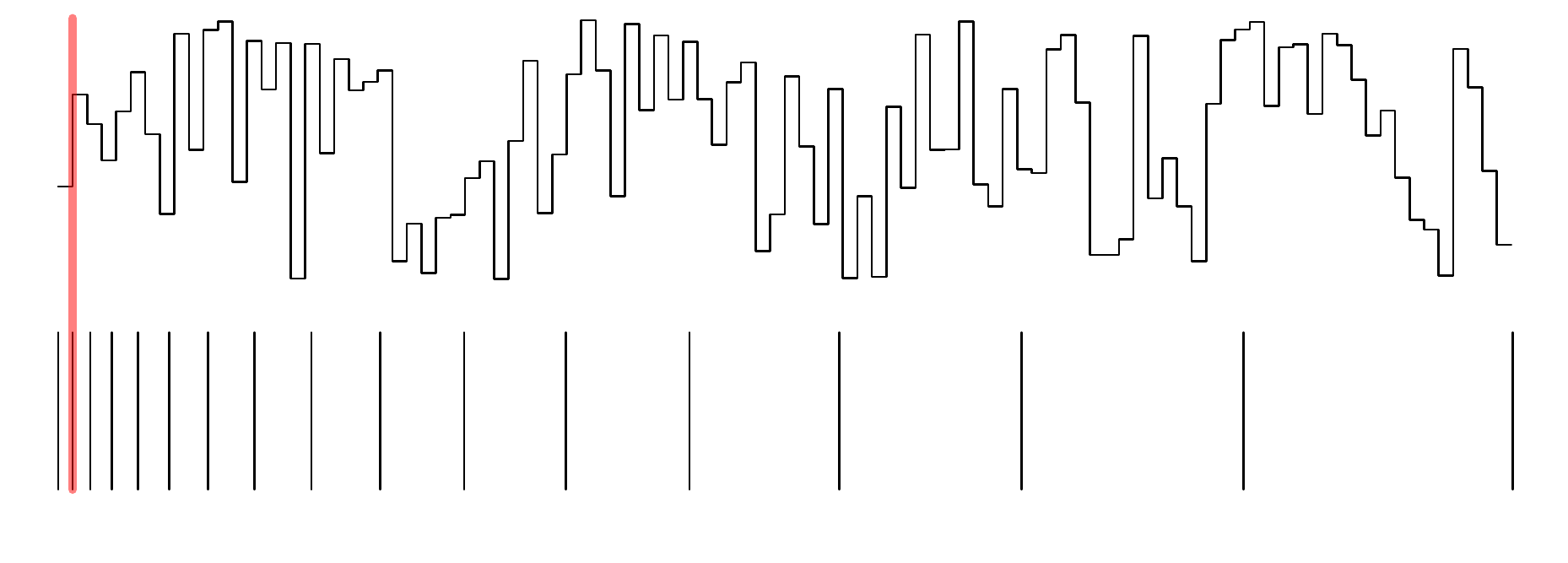}\\
		\textbf{c}\\
			&\multicolumn{3}{c}{\includegraphics[width=0.9\columnwidth]{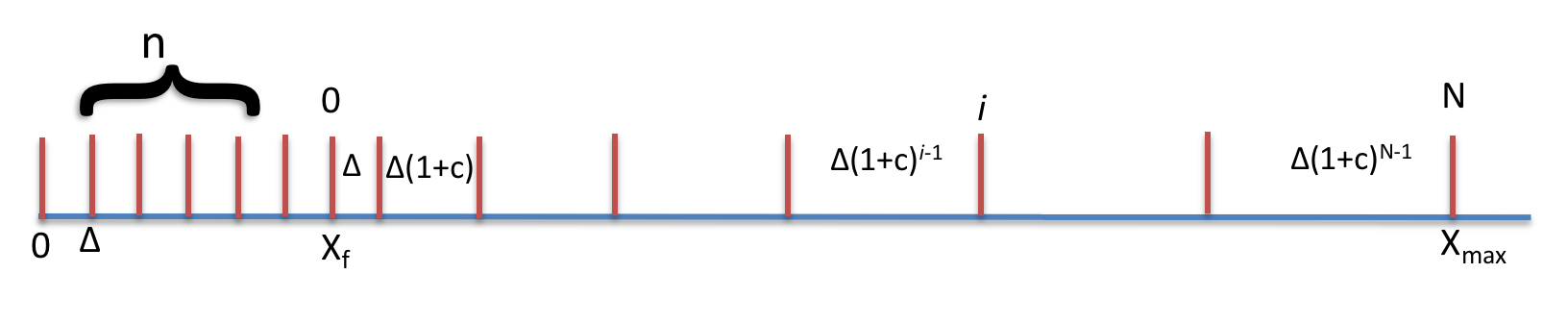}}
	\end{tabular}
	\end{center}
	 \caption{
			\textbf{Rationale for inclusion of a fovea.}
			 \textbf{a-b.} 
			Functions are shown with different scales, as in
					Figure~\ref{fig:adaptive} for logarithmically-spaced
					receptor locations.  \textbf{a.} With a function
					scale that is much larger than $\Deltamin$, several values
					of the function can be represented (here $I=5$).
					\textbf{b.} When the scale goes to $\Deltamin$, then much
					less total information can be represented.  Here only one
					function value fits into the veridical region.  Note that
					the information conveyed by the same set of receptors
					decreases as $s$ decreases, as in the dashed lines in
					Figure~\ref{fig:informationc}.
		   			\textbf{c.} Schematic for notation for hybrid neural scaling.  
		 The first part of the scale, analogous to a fovea, has $n$ evenly
				 spaced receptors from $0$ to $\xmin$.
			   The second part of the scale region, from $\xmin$ to $\xmax$
	   has $N$ receptors spaced such that the ratio of adjacent differences is
	   $1+c$.   The derivation suggests $n$ should be fixed at $1/c$, whereas
	   $N$ can in principle grow without bound.  If $\xmin=\Deltamin/c$, the
	   scale starts precisely at $x=0$.
	   \label{fig:schematicfovea}
	 }
\end{figure}

If we set $c>0$ such that all receptors are placed in constant ratio 
the responsiveness to functions of different scales is very different.
Let us place the zeroth receptor $x_0$ at $\xmin$, and
the first receptor at $x_1 = \xmin + \Deltamin$.  Continuing with $\Delta_i =
\left(1+c\right) \Delta_{i-1}$, we find that the spacing of the $i$th receptor
is given by
\begin{equation}
	\Delta_i 
   	= 
   	\Deltamin \left(1+c\right)^{i-1}
	\label{eq:deltai}
\end{equation}
The position of the $i$th receptor is thus given by the geometric series
\begin{eqnarray}
	x_i &=&  \xmin + \frac{\Deltamin}{c}\left[\left(1+c\right)^{i} -1
	\right] \label{eq:xi}
\end{eqnarray}
Equations~\ref{eq:deltai}~and~\ref{eq:xi} show that when $c>0$ the
range of $x$ values and scales that can be represented with $N$ receptors goes
up exponentially like $(1+c)^N$. 
For a function of scale $s$, the set of ratio-scaled receptors has a
veridical region ranging from $\xmin$ to 
\begin{equation}
	\xcrit = \left\{ \begin{array}{lr}
		 s\frac{1+c}{c} - \frac{\Deltamin}{c} + \xmin &  s \leq \Delta_N  \\
				 \\
		\xmax  &  s > \Delta_N
		\end{array}
	\right.
\end{equation}
The amount of information present in the function in the veridical region is
just $(\xcrit-\xmin)/s$, which is given by
\begin{equation}
	I_{c>0} = \left\{ \begin{array}{lr}
		 \frac{1+c}{c} - \frac{\Deltamin}{sc}&  s \leq \Delta_N  \\
				 \\
		\frac{\xmax-\xmin}{s} &  s > \Delta_N
		\end{array}
	\right.
	\label{eq:Icgreater0}
\end{equation}
The second expression, which happens when $s$ is larger
than the largest
spacing among the set of receptors, is closely analogous to the case with
constant spacing ($c=0$), decreasing like $s^{-1}$.  However, the first expression
with $s \leq \Delta_N$ includes a term $(1+c)/c$ which is independent of $s$.  The second term is small
when $s$ is large, so that for large values of $s$ the information conveyed is
approximately constant.  However, when $s$ is small the second term is
substantial and constant ratio spacing ($c>0$) fails to convey much
information about the function in the veridical region. 
When $s=\Deltamin$ the veridical region includes only one value of the
function (see Figure~\ref{fig:schematicfovea}a).  The dashed line
in Figure~\ref{fig:informationc} shows the analytic result when $c > 0$ and is
constant across all receptors.  


\subsection{A ``fovea'' surrounded by a logarithmic scale}
Note that if the $\frac{\Deltamin}{sc}$ term in Eq.~\ref{eq:Icgreater0} were
canceled out, the amount of information in the veridical region conveyed by
the set of receptors would be precisely invariant with respect to $s$ over the
central range.  This can be accomplished by 
preceding the region of ratio spacing ($c>0$)
with a region of constant spacing ($c=0$).   In order to equalize the
information carried at small scales, the region of constant spacing should
have $n = \frac{1}{c}$ receptors each spaced by $\Deltamin$, as in
Figure~\ref{fig:schematicfovea}b.
Adding a set of constantly spaced receptors enables constant information
transfer over a range of scales starting exactly at $\Deltamin$ (flat solid
curves in Figure~\ref{fig:informationc}).  
When the region with ratio
spacing starts at $\xmin=\frac{\Deltamin}{c}$, then the region of constant spacing
starts at $x=0$.  Note that the number of receptors in the fovea is controlled
only by $c$ and is independent of the number of receptors in the region with
ratio spacing.  The ratio of adjacent receptor spacings is constant within
each region and only fails to hold exactly at the transition between regions. 


\section{Discussion}

Current technology is rapidly leading to a situation where we can design
intelligent agents.  The results in this paper point to a design choice that
may have played out on an evolutionary time scale.  To the extent
neural representations of very different one-dimensional quantities all obey
the same scaling laws, it suggests convergent evolution optimizing
some design principle.  One strategy that could be taken in designing receptor
systems is to optimally adapt the organism to the behaviorally-relevant
statistics of the environment.  Evolution has certainly made this choice in a
number of cases---the putative ``fly detectors'' in the frog's visual system
are a famous example of this approach (\citeNP{LettEtal59,MatuEtal60}, see
also \citeNP{HerzBarl92}).  However, optimizing receptor arrays to a specific
configuration of the world comes with a cost in terms of flexibility in
responding to changes in the world.  

\subsection{Logarithmic receptor scales and natural statistics}
In this paper, we have pursued the implications of a neural uncertainty
principle---maintaining maximum ignorance about the statistics of the
world---for the design of sensory receptors.
Operationally, this means we have made only minimal assumptions about the
statistics of the functions to be represented, limiting ourselves to uniform
priors and the assumption of independence.  Had we assumed a non-uniform
distribution for the prior of $s$, this would have required us to estimate at
least one parameter.  Similarly, if the successive values of $s_x$ were not
independent, we would have to have estimated at least one parameter to
characterize the nature of the dependence.  It can be shown that
the arguments developed in the first section of this paper (``Optimal receptor
distribution \ldots'') for uniform priors also generalize
to scale-free (power law) priors (see footnote~\ref{foot:power}).  Much empirical work has characterized
statistical regularities in the world. For instance, power spectra in natural
images tend to be distributed as a power law with exponent near $-2$
\cite{Fiel87,RudeBial94,SimoOlsh01}.  Similarly, it has been argued that power
spectra for auditory stimuli are distributed as a power law with exponent $-1$
\cite{VossClar75}.   Perhaps it would be adaptive to design a set of receptors
that is optimized for these naturally occurring statistics
\cite<e.g.,>{WeiStoc12,Pian16}.

Results showing power law spectra with similar exponents for natural images
mask variability across orientations and categories of images
\cite{TorrOliv03}.  For instance, the exponent observed,  and especially
second-order statistics, vary widely across pictures of landscapes \emph{vs}
pictures of office environments or pictures of roads.  To the extent these
statistics differ, if receptors were optimized for any one of these categories
of images, they would be suboptimally configured for other environments
\cite{WeiStoc12}.  Moreover, because the eyes are constantly in motion, the
statistics of natural images are not necessarily a good proxy for the
statistics of light landing on the retina averaged by synaptic time constants.
More concretely, the temporal variation due to fixational eye movements has
the effect of whitening images with conventional power law spectra
\cite{KuanEtal12,Rucc08}.   In the time domain, it can be shown that in the
presence of long-range correlated signals, logarithmic receptor spacing is
optimal for predicting the next stimulus that will be presented
\cite{ShanHowa13}.    

A derivation based on a uniform prior is more general than a belief that
the statistics of the world should be power law.  The former would apply
across a range of environments and equally well apply to any world with some
unknown statistics.  Moreover, the development in the next section (entitled
``Global function representation''), which predicts the existence of a fovea
under some circumstances, would be quite different if we had a strong prior
belief about the probability of observing a particular scale $s$.  
If the goal was to maximize the information across states of the world, and if
there was a non-uniform prior about the distribution of scales, we would not
have obtained asymptotically logarithmic receptor spacing.

\subsection{Specific predictions deriving from the approach in this paper}
There have been a great many other approaches to understanding the 
ubiquity of logarithmic psychological scales  and the more general problem of
constructing psychological scales with well-behaved mathematical properties
from continua in the world.  These are briefly reviewed in the next subsection
(entitled ``Placing this work in historical context'') with special attention
to drawing contrasts with the present paper.  However, to our knowledge, the
predictions about the scaling of the foveal region are unique to the present
approach.

Asymptotically, a logarithmic neural scale enables a set of receptors to provide
equivalent information about functions of a wide range of intrinsic scales,
implementing the principle of neural equanimity.  However, logarithmic neural
scales are  untenable as $\Delta$ goes to zero---the number of receptors
necessary tends to infinity and the set of receptors conveys less information
about functions with scale near the smallest receptor spacing (corresponding
to the maximum resolution).  In the absence of a fovea, there is a small-scale
correction (dashed lines Fig.~\ref{fig:informationc}).  A region of constant
receptor spacing near zero---a fovea---solves this problem allowing the set of
receptors to carry the same amount of non-redundant information about
functions with every possible scale ranging from $\Deltamin$ to $\Delta_N$.  

Equalizing the information across scales results in a fixed number of
receptors within the fovea.  Critically, the number of receptors along a
radius of the fovea, measured in units of $\Deltamin$, depends only on $c$.
The value of $c$ can be estimated from noting the slope of the
line relating receptor size to the center of the receptive field, as in
Figure~\ref{fig:FreeSimo}b.
\footnote{This prediction  is at least roughly consistent with the trend of
		the hinged linear functions generated by \cite{FreeSimo11} shown in
		Fig.~\ref{fig:FreeSimo}b, but this visual
impression should not be taken as strong quantitative evidence.}
Note that the value of $c$ is estimated from receptive fields outside of the
fovea, whereas $\Deltamin$ and the number of receptors along a radius are
estimated from information about the fovea.  As such, measurement of the two
quantities ought to be completely independent.  This quantitative relationship
constitutes a specific prediction of this approach and can be evaluated across
brain regions within the same modality, and even across modalities.

\subsection{Placing this work in historical context}

The present paper provides a rational basis for logarithmic neural scales and
the presence of a ``fovea'' for values of $x$ near zero.  It also adds to a
long tradition of work in mathematical psychology organized along several
themes.

\subsubsection{Measurement theory}

Researchers in mathematical psychology have long considered the form of
psychological spaces.  This work has concluded, \emph{contra} the present
approach,  that there is not a privileged status for logarithmic psychological
scales.  Fechner's \citeyear{Fech60} reasoning for logarithmic psychological
scales started with the empirical observation of a constant Weber fraction and
then made what was apparently a straightforward conclusion: integration of the
Weber law results in logarithmic psychological scale.  That is, the Weber law
states that the change in the psychological discriminability $\Delta p$ due to
a change in the magnitude of a physical stimulus $\Delta x$ goes like $\Delta
p  \propto \frac{\Delta x}{x}$.  Taking the limit as $\Delta x$ goes to zero
and integrating gives a logarithmic scale $p = \log x + C$.  On its face this
seems reasonable; surely the psychological distance between two stimuli should
be the sum of the JNDs along the path between them.  

\citeA{LuceEdwa58} noted  that Fechner's reasoning is not in general sound;
the integration procedure is only valid if the empirical Weber fraction result
holds. \citeA{Luce59} showed that a logarithmic psychological scale is one of
a class of relationships that can map a physical scale with a natural zero (a
ratio scale) onto an interval scale that is translation-invariant
\cite{Stev46}.  \citeA{DzhaColo99,DzhaColo01} developed a much more general
framework for constructing multidimensional psychological spaces from local
discriminability functions (see \citeNP{LuceSupp02} for an accessible
introduction to the history of these questions) that also does not find any
privileged status for logarithmic functions.  And none of these approaches
provides a natural account for the existence of a ``fovea''---a region of
heightened discriminabiligy near $x=0$.  To the extent that logarithmic neural
scales are a general property of the brain, the neural considerations
described in this paper provide a potentially important complement to
measurement theory.

\subsubsection{Recent approaches to Weber-Fechner scaling}
More recently, investigators in mathematical psychology and related fields
have also considered the rationale underlying the apparent ubiquity of the
Weber-Fechner law.  These approaches have in general taken a more restrictive
approach than the quite general considerations in this paper.  Moreover, they
do not lead to the specific predictions regarding the fovea derived here.
Some recent approaches have noted that if the psychological scale is designed
to minimize a relative error measure \cite{SunEtal12,PortSvai11}, logarithmic
scales naturally result.   For instance, in stimulus quantization, the goal is
to choose a discrete quantization of the stimulus space in order to minimize
the expected value of some measure of reconstruction error between the true
stimulus and the quantized stimulus.  If one minimizes mean squared error, the
optimal quantization is uniform.  However, if one attempts to minimize
relative error, and if the quantization is constrained to have a fixed
entropy, it can be shown that the optimal quantization is on a logarithmic
scale independent of the input stimulus statistics \cite{SunGoya11,SunEtal12}.
\citeA{Wilk15} compared the suitability of various relative error
measures from an evolutionary perspective.  These approaches depend critically
on the assumption of relative error measures, without explaining why those
might be desirable (other than that they result in Weber-Fechner scales).  

Other recent approaches that result in Weber-Fechner spacing make specific
assumption about the statistics of the world.  For instance \citeA{ShouEtal13}
described the variability in perception as due to the variability in spike
count statistics from a neuron with some I/O function and Poisson variability.
They concluded that the Weber law is optimal if the world has power law
statistics.  Similarly, \citeA{Pian16} derived a Weber-Fechner scale for
numerosity under the assumption that the probability that each possible number
will be utilized goes down like a power law.  This builds on earlier work
deriving a rational basis for power law forgetting on the probability of use
of a memory a certain time in the past \cite{AndeScho91}.  \citeA{WeiStoc12}
argued that receptor spacings should be constructed so that each receptor
carries the same amount of information about the world.  This depends on the
statistics of the world and a logarithmic neural scale results only if the
world has power law statistics with exponent $-1$.  To the extent the world
has different statistics in different modalities and different environments,
these approaches are limited in accounting for the ubiquity of Weber-Fechner
neural scales.   The approach in the present paper is more general in that it
does not make any strong prior assumptions about the statistics of the world.

\subsubsection{Universal exponential generalization}
On its face, logarithmic neural scales seem to be closely related to
Shepard's work on  universal exponential generalization \cite{Shep87}.  If the
mapping between the physical world and the neural scale is logarithmic, one
would expect that the mapping between the neural scale and the physical world
is exponential.  Indeed, the approach in the section ``Optimal receptor
distribution \ldots'' was very much inspired by Shepard's pioneering approach
to the structure of an abstract psychological space \cite{Shep87}.    However,
the results are actually quite distinct.

\citeA{Shep87} studied the problem of generalization; if a particular stimulus $x$
leads to some response, what is the optimal way to generalize the response to
other stimuli in the neighborhood of $x$?  Shepard's derivation hypothesizes a
consequential region of unknown size.  Although the results depend on one's
prior belief about the distribution of sizes, if one uses a Copernican
principle like Gott to set the prior, one obtains an exponential
generalization gradient.  The size of the consequential region in Shepard's
work plays a role analogous to the scale of the function in this paper.  And
an exponential gradient is analogous to a logarithmic scale.   

The value $1+c$ serves the role of the base of the logarithm in the scales
derived here.  This corresponds (roughly) to an exponential generalization
gradient that goes like $(1+c)^x$. In Shepard's framework, the ``space 
constant'' of the exponential gradient is controlled by the expected size of
the consequential region, which functions like a prior choice of scale.  Note
that the neural scales developed here would not lead to an exponential
generalization gradient around any particular point $x$.  The width of
receptive fields (which is proportional to $\Delta$) provides a natural lower
limit to generalization.  We would not expect this generalization gradient to
be symmetric in $x$ (see Fig.~\ref{fig:FreeSimo}a).  Of course psychological
generalization need not be solely a function of neural similarity and is  more
difficult to observe than receptive fields. 

\subsection{Neural scales for cognitive dimensions}
The quantitative evidence for spacing of visual receptive fields is quite
strong, due to intense empirical study for decades
\cite{DaniWhit61,HubeWies74,VanEEtal84}.  The derivation in this paper applies
equally well to any one-dimensional quantity over which we want to represent a
function.  If the neural uncertainty principle is a design goal for the
nervous system, it should be possible to observe similar scaling laws for
representations of other dimensions.  As discussed in the introduction, there
is evidence that the brain maintains receptive fields for dimensions that do
not correspond to sensory continua.  In the visual system, a neuron's spatial
receptive field is the contiguous region over visual space that causes the
neuron to be activated.  Cognitive continua can also show the same coding
scheme.  For instance, in a working memory task that required macaques to
remember the number of stimuli presented over the delay, neurons in lateral
prefrontal cortex responded to a circumscribed set of numbers to be
remembered.  That is, one neuron might be activated when the number of stimuli
to be remembered is 3-5, but another neuron might be activated when the number
to be remembered is 4-7.

Time cells fire during a circumscribed period of time within a delay interval.
Each time cell can be thought of as having a receptive field over past time.
In much the same way that a neuron with a visually-sensitive receptive field
fires when a stimulus is in the appropriate part of the visual field, so too
the time cell fires when a relevant stimulus, here the beginning of the delay
interval, enters its temporal receptive field.  Because different time cells
have different receptive fields, as the stimulus recedes into the past with
the passage of time, a sequence of time cells fire.  In this way, receptive
fields tile time, a continuous dimension in much the same way that
eccentricity or numerosity would be coded.

Time and numerosity do not correspond to sensory receptors in the same way
that, say, eccentricity in the visual system does. Nonetheless, the logic of
the arguments in this paper apply equally well to time and numerosity.  Do
these ``cognitive'' receptive fields show logarithmic spacing?  Although this
is still an open empirical question, qualitative findings are consistent with
logarithmic spacing.  Logarithmic spacing would imply that the width of time
fields should increase linearly with the time of peak firing within the delay.
Although a precisely linear relationship has not been established, it is
certain that the width of time fields increases with time of peak firing.
This is true in the hippocampus \cite{MacDEtal11,SalzEtal16}, entorhinal
cortex \cite{KrauEtal15}, striatum \cite{AkhlEtal16,MellEtal15,JinEtal09}, and medial
prefrontal cortex \cite{TigaEtal16}.  Similarly, logarithmic spacing implies
that the number density of cells with a time field centered on time $\tau$
should go down like $\tau^{-1}$.  While this quantitative relationship has not
been established thus far, it is certain that the number density goes down
with $\tau$.   This qualitative pattern at least holds in all of the regions
where time cells have been identified thus far.

In addition to logarithmic spacing, the logic motivating the existence of a
fovea would apply equally well to cognitive dimensions.  This does not imply
that foveal organization ought to be ubiquitous.  We would expect the size of
the  fovea to be large when $c$ is small.  The cost of choosing a small value
of $c$ is that the range of scales that can be represented with a fixed $N$
goes down dramatically (Eq.~\ref{eq:deltai}).  In domains where there is no
natural upper limit to $x$, such as numerosity or (arguably) time, this
concern may be quite serious.  However, the cost of failing to represent
arbitrarily large scales is perhaps not so high in cases where there is a
natural upper bound on the function to be represented. In the case of vision,
there is a natural upper bound to the value of eccentricity that could be
observed.  In the case of functions of numerosity, the subitizing range, the
finding that small integers are represented with little error, may be
analogous to the fovea.  In the case of functions of time, the analog of a
fovea would correspond to something like traditional notions of short-term
memory. In any case, direct measurement of $c$, which can be estimated from
the width of receptive fields as a function of their peak, would provide
strong constraints on whether or not there ought to be a ``fovea'' for
numerosity and/or time.

\subsection{Constructing ratio scales for cognitive dimensions}  Ultimately,
		one can understand the spacing of the retinal coordinate system, or
		other sensory domains, as the result of a developmental process that
		aligns receptors along the sensory organ.  However, if these arguments
		apply as well to non-sensory dimensions such as number and time, then
		this naturally raises the question of how the brain constructs
		``receptors'' for these entities.  One recent hypothesis for
		constructing representations of time, space and number describes
		time cells---which have receptive fields for particular events in the
		past---as extracted from exponentially-decaying neurons with long
		time constants.  The requirement for logarithmic neural scales amounts
		to the requirement that the time constants of exponentially-decaying
		cells should be organized such that ``adjacent'' cells, along some
		gradient, should have time constants in a constant ratio.
		
		There is now good evidence for long time constants in
		cortex both \emph{in vitro}  \cite{EgorEtal02} and \emph{in vivo}
		\cite{LeitEtal16}.  The long time constants can be implemented at the
		single cell level using known biophysical properties of cortical
		neurons \cite{TigaEtal15}.
		Noting that a set of exponentially-decaying cells
		encode the Laplace transform of the history, it has been proposed that
		time cells result from an approximate inversion of the Laplace
		transform \cite{ShanHowa13}.  The inversion can be accomplished with,
		essentially, feedforward on-center/off-surround receptive fields.  
		
		The mathematical basis of this approach is extremely powerful.  It is
		straightfoward to show that the computational framework for a
		representation of time can be generalized to space and number
		\cite{HowaEtal14} or any variable for which the time
		derivative can be computed.  Moreover, access to the Laplace domain
		means one can implement various computations, for instance
		implementing translation \cite{ShanEtal16} or comparison operators
		\cite{HowaEtal15}.  To the extent that representations from different
		domains utilize the same scaling relationships, access to a set of
		canonical computations could lead to a general computational framework
		for cognition.

\bibliography{/Users/marc/doc/bibdesk}

\end{document}